\documentclass[a4paper]{article}

\usepackage{a4wide}

\usepackage{amsmath}%
\usepackage{amsfonts}%
\usepackage{amssymb}%
\usepackage{graphicx}
\usepackage{cite}

\usepackage{bm}

\begin{document}

\title{Transformation Acoustics in Generic Elastic Media}
\date{October 25, 2012}
\author{Luzi Bergamin\thanks{luzi.bergamin@kbp.ch}\\ KBP GmbH, Fliederweg 10, 3007 Bern, Switzerland}

\maketitle

\begin{abstract}
 In this work a transformation acoustics scheme for generic elastic media is developed. Our approach starts form the decomposition of the elasticity tensor in terms of its eigentensors, an idea previously used by Norris. While Norris' transformation acoustics is restricted to the special class of so-called pentamode materials, we show that a similar scheme can be defined for the most general elasticity tensor. As in case of Norris' model (and in sharp contrast to transformation optics), the compatibility equations of the transformation medium are not purely algebraic and it is not guaranteed that solutions to these equations exist for any choice of material parameters and coordinate transformation. Nonetheless, it is shown that our scheme yields new cloaking solutions for certain classes of materials. In particular, we present the first application of a transformation based device for a non-scalar wave equation outside of the field of electromagnetics.
\end{abstract}

\section{Introduction}
Transformation optics \cite{Pendry:2006Sc,Leonhardt:2006Sc,Leonhardt:2006Nj} is one of the most important design tools in the field of artificial electromagnetic materials (metamaterials). A unique combination of characteristics makes this tool extremely powerful: the trajectories of light in the metamaterial are  simply defined as geometric deformations (coordinate transformations) of the trajectories in free space; the solutions of the Maxwell equations in the metamaterial together with its material properties are obtained from the free space solutions and the deformations in a completely algebraic way; once certain intuitive restrictions on the deformations are respected, all interfaces between free space and the metamatrial are automatically reflectionless \cite{Bergamin:2009In}.

This combination for the first time allowed to derive the necessary material parameters of almost arbitrary optical devices in a simple, intuitive way. Not surprisingly, besides the original proposal of a cloak \cite{Pendry:2006Sc,Leonhardt:2006Sc} a pletora of new concepts was proposed (see, for example, \cite{Leonhardt:2006Nj,Rahm:2008Pl,Li:2008Cc,Chen:2010Rv}).

Of course, these successes raised the question, whether similar tools can be defined for other theories of physics, notably different types of waves. The first attempt to apply similar concepts to acoustic waves \cite{Milton:2006Ac} appeared in the same year as the electromagnetic cloak and became an independent field of research \cite{Greenleaf:2008Qa,Norris:2008Ps,Cummer:2007Pa,Chen:2010Ra}. Still, the manipulation of acoustic waves turns out to be much harder than of its electromagnetic counterpart. The reason for this lies in a very specific characteristic of the Maxwell equations, which is not shared by most other wave problems in physics: the existence of a premetric formulation. Indeed, by a suitable choice of electromagnetic variables, the complete Maxwell equations can be written in a form that does not include any spacetime metric or any object derived from the spacetime metric \cite{Hehl:2003}. Therefore, in these variables the Maxwell equations --- and in consequence any solution thereof --- do not carry any information about the local structure of spacetime or the local choice of coordinates\footnote{This fact is often referred to as form invariance of the Maxwell equations. This term can be misleading, since --- in a suitably covariant notation --- any sensible law of physics is form invariant under diffeomorphisms.}. This peculiar feature of electrodynamics makes transformation optics possible: starting with a free space solution of the Maxwell equations in coordinates $\bm x$ and with spacetime metric $\bm g$, one applies a coordinate transformation to obtain the same solution in coordinates $\bar{\bm x}$ with spacetime metric $\bar{\bm g}$; since the Maxwell equations do not depend on the metric, one can easily interpret the transformed solution as a solution in the original coordinates $\bm x$ with spacetime metric $\bm g$, which is exactly the transformation optics solution that belongs to the coordinate transformation $\bm x \rightarrow \bar{\bm x}(x)$. Though mostly used in the context of deformations of the free space solutions, transformation optics can be applied to any constitutive law, including nonlinear \cite{Bergamin:2011Nl,Paul:2012Oe} and nonlocal \cite{Castaldi:2012Nl} media relations.

The theory of acoustic waves in elastic bodies is not premetric and thus the scheme sketched above cannot apply to this theory in general. Still, it was found \cite{Greenleaf:2008Qa,Greenleaf:2009Am} that in certain special cases transformation acoustics exists. Consider a limit of the full theory that can be simplified to a scalar field $\phi(\bm x,t)$ obeying a Helmholtz equation
\begin{equation}
\label{eq:intro1}
\bm \Delta \phi = \ddot \phi\ .
\end{equation}
In a covariant notation this equation reads
\begin{equation}
D_i \gamma^{ij} D_j \phi = \ddot \phi\ ,
\end{equation}
where $\bm D$ is the covariant derivative with respect to the spatial metric $\bm \gamma$. The second covariant derivative just acts on a scalar and thus simply can be replaced by a plain derivative $\bm \partial$. The first covariant derivative constitutes the gradient of a vector
\begin{equation}
 D_i v^i\ , \qquad v^i = \gamma^{ij} \partial_j \phi\ ,
\end{equation}
which can be written as
\begin{equation}
\label{eq:intro2}
 D_i v^i = \frac{1}{\sqrt{\gamma}} \partial _i \left( \sqrt{\gamma} v^i\right)\ ,
\end{equation}
with $\gamma = \det(\gamma_{ij})$. Therefore, the Helmholtz equation becomes
\begin{equation}
\label{eq:intro3}
\frac{1}{\sqrt{\gamma}} \partial _i \left( \sqrt{\gamma} \gamma^{ij} \partial _j \phi \right) = \kappa \partial_i M^{ij} \partial_j \phi = \ddot \phi\ .
\end{equation}
This equation appears to be metric independent if $\kappa$ and $M^{ij}$ can be interpreted in a suitable way as material parameters. It is important to realize that $\kappa$ does not transform as a scalar, but as a relative scalar and, similarly, $\bm M$ is a tensor density rather than a tensor. To our knowledge, all transformational design tools outside of transformation optics known so far are based on Eq.~\eqref{eq:intro3}. Consequently, all these design tools are restricted to scalar wave problems. It is the purpose of this paper to suggest a route to overcome this restriction, whereby linear elastodynamics will serve as a guiding example.

The idea presented here is a generalization of Norris' pentamode transformation acoustics \cite{Norris:2008Ps}. In this model, Norris' defines a special class of metamaterials in elastodynamics, whose wave equation essentailly reduces to Eqs.\ \eqref{eq:intro1}--\eqref{eq:intro3}. After a brief review of linear elastodynamics (Sec.\ \ref{se:elasto}) and the special case of pentamode materials (Sec.\ \ref{se:norris}) we will show in Sec.\ \ref{se:GenericMedia} that the wave equation of a large class of elastic media can be cast into the form
\begin{equation}
\label{eq:intro4}
\sum_J K_I D_i G_{IJ}^{ij} D_j p_J = \ddot p_I\ ,
\end{equation}
where the indices $I,J$ take values $(1,2,\ldots,6)$. This means, that the tensorial wave equation of elastodynamics in these cases can be rewritten as a system of six coupled scalar wave equations. In consequence, the transformation acoustics scheme based on Eq.\ \eqref{eq:intro3} conceptually can be applied to this large class of media straightforwardly, but the ensuing compatibility equations for the material parameters turn out to be complicated and --- as already found by Norris \cite{Norris:2008Ps} --- non-algebraic. The concept of Eq.\ \eqref{eq:intro4} can be further generalized to include all linear elastic media by adding appropriate source terms to that equation. Despite their complexity it is shown in Sec.\ \ref{se:examples} that new solutions for linear radial transformations in cylindrical and spherical coordinates can be found based on our concept, which includes new types of cloak solutions.

\section{Linear elastodynamics}
\label{se:elasto}
Our model are the linearized elasticity equations \cite{Amenzade:1979Mi,Marsden:1994Do} in terms of the stress tensor $\bm \sigma = (\sigma^{ij})$, the infinitesimal strain tensor $\bm e = (e_{ij})$ and the elasticity tensor $\bm C = (C^{ijkl})$. In order to fix our notation we provide a brief overview of the relevant equations in this section.

The stress tensor is subject to momentum and angular momentum balance. Due to the latter, $\bm \sigma$ must be symmetric, $\sigma^{ij} = \sigma^{ji}$, while the former can be cast into the form
\begin{equation}
D_j \sigma^{ij} + \rho F^i=\rho W^i\ . 
\label{eq:intr1}
\end{equation}
Here, $\rho(x)$ is the density of the volume element, $\rho F^i$ are the total volume forces and $W^i$ is the acceleration of the volume element. In this paper we always assume $\bm F \equiv 0$. Starting from Eq.\ \eqref{eq:intr1} leads to a very restricted version of transformation acoustics, in particular, the special case of inertial transformation acoustics \cite{Cummer:2007Pa,Chen:2007Al,Cummer:2008Pl} is not covered at all. To get more flexibility, a anisotropic mass density is indispensable (see Ref.\ \cite{Norris:2008Ps} and references therein). Therefore, for the purpose of transformation acoustics Eq.\ \eqref{eq:intr1} is replaced by
\begin{equation}
D_j \sigma^{ij}=\rho^i{}_j W^j\ . 
\label{eq:intr1.1}
\end{equation}

The infinitesimal (or small) strain tensor is the Lie derivative along the displacement vector $\bm u$ of the spatial metric,
\begin{align}
\bm e &= \frac12 \mathcal L_u \bm \gamma\ , & e_{ij} &= \frac12 \left(D_i u_j + D_j u_i\right)\ ,
\label{eq:intr2}
\end{align}
where $\bm D$ is the covariant derivative with respect to the spatial metric $\bm \gamma$. The elasticity tensor defines a linear constitutive law between the stress and the strain tensor:
\begin{equation}
\sigma^{ij} = C^{ijkl} e_{kl}
\label{eq:intr3}
\end{equation}
Finally, the acceleration vector $\bm W$ must be related to the displacement vector field by
\begin{equation}
\bm W = \frac{d^2 \bm u}{dt^2} = \ddot{\bm u}\ .
\label{eq:intr4}
\end{equation}

Due to the characteristics of the stress and strain tensor $\bm C$ must be symmetric in its first two and in its last two indices. Thus, $\bm C$ has 36 independent parameters and may be represented as a generic $6\times6$ matrix. Additional constraints on the elasticity parameters emerge if an elastic potential W with
\begin{align}
\sigma^{ij} &= \frac{\partial W}{\partial e_{ij}}\ , & C^{ijkl} &= \frac{\partial^2 W}{\partial e_{ij}\partial e_{kl}}\ , 
\label{eq:decomp1}
\end{align}
shall be defined. In that case $C^{ijkl} = C^{klij}$ and in analogy to electromagnetism \cite{Hehl:2003} such a material can be called skewonless. As already mentioned by Milton and Cherkaev \cite{Milton:1995Et}, the elasticity tensor of any skewonless material can be decomposed as\footnote{In contrast to indices of physical space, no summation convention is invoked for the space of eigentensors and eigenvectors, labeled by capital Latin indices.}
\begin{equation}
C^{ijkl} = \sum_{I=1}^6 K_I S_I^{ij} S_I^{kl}
\label{eq:decomp}
\end{equation}
This decomposition is not unique. The scalars $K_I$ can be chosen as eigenvalues to the normalized eigentensors $S^{ij}$, $\mbox{Tr}( \bm S \cdot \bm S) = 1$, alternatively they can be fixed as numbers $K_I = \pm1, 0$. In general, the $K_I$ are scalar fields. In contrast to Ref.~\cite{Milton:1995Et} the decomposition is not fixed in this article.

\section{Norris' transformation acoustics revisited}
\label{se:norris}
An important precursor to results derived in this work is the paper by Norris \cite{Norris:2008Ps}. Thus, it appears helpful to reformulate the results of that paper in our notation in a first step. A more careful re-derivation following the logic and notation of this paper can be found in Ref.\ \cite{Favaro2012:Ad}. 

Norris considers a very special material, a so-called pentamode medium. In terms of the decompositon \eqref{eq:decomp} this class of materials is characterized by the fact that it has only one nonvanishing eigenvalue:
\begin{align}
K_1 &= K & K_2 &= K_3 = K_4 = K_5 = K_6= 0
\label{eq:norris1}
\end{align}
This is also referred to as a material with five ``easy modes'', since five out of six modes of the stress tensor are not related to the modes of the elasticity tensor.

In addition, Norris requires that the eigentensor $\bm S$ which corresponds to the nonvanishing eigenvalue $K$ must have vanishing divergence:
\begin{equation}
D_i S^{ij} = 0
\label{eq:norris2}
\end{equation}
As shown by Norris, the wave equation simplifies in a very interesting way for this special system. A quantity with dimension of a pressure can be defined by contracting $\bm S$ with the strain tensor $\bm e$:
\begin{equation}
p = - K S^{ij}e_{ij} = -K S^{ij} D_i u_j\ .
\label{eq:norris3}
\end{equation}
With this definition, the stress tensor becomes $\sigma^{ij} = -S^{ij} p$ and, consequently, the differential equation of the stress tensor \eqref{eq:intr1.1} can be written in the compact form
\begin{equation}
D_j \sigma^{ij} = - S^{ij} D_j p = \rho^{i}{}_j \ddot u^j\ .
\label{eq:norris4}
\end{equation}
The time derivative of the displacement field can be related to the time derivative of $p$ by differentiating Eq.\ \eqref{eq:norris3} with respect to time
\begin{equation}
\ddot p = - K S^{ij} D_i \ddot u_j\ ,
\label{eq:norris5}
\end{equation}
where we assumed that the elasticity tensor is not time dependent. Combining \eqref{eq:norris4} and \eqref{eq:norris5} one arrives at a wave equation for $p$:
\begin{equation}
\ddot p = K D_i \left(S^{ij} \rho^{-1}_{jk} S^{kl}\right) D_l p = K D_i G^{ij} D_j p
\label{eq:norris6}
\end{equation}
It is seen that for the special case of a pentamode medium the wave equation of elasticity theory reduces to a scalar wave equation of the ``pseudo-pressure'' $p$. The tensor $\bm G = \bm S \cdot \bm \rho^{-1} \cdot \bm S$ can be interpreted as effective metric, but it is important to realize that the covariant derivatives are still defined with respect to the spatial metric $\bm \gamma$ rather than $\bm G$. Turning the covariant derivatives into plain derivatives (cf.\ Eqs.\ \eqref{eq:intro1}--\eqref{eq:intro3}) yields
\begin{equation}
\ddot p = \frac{K}{\sqrt{\gamma}} \partial_i \sqrt{\gamma} G^{ij} \partial_j p\ . 
\label{eq:norris6.2}
\end{equation}

Consider a coordinate transformation $\bm x \Longrightarrow \bar{\bm x}(x)$ applied on the wave equation \eqref{eq:norris6}. Since we use a covariant formulation the effect of the transformation boils down to putting a bar over each non-scalar quantity:
\begin{equation}
\ddot{p} = K \bar D_i \left(\bar S^{ik} \bar \rho^{-1}_{kl} \bar S^{lj}\right) \bar D_j p = K \bar D_i \bar G^{ij} \bar D_j p\ ,
\label{eq:norris8}
\end{equation}
which in the language of Eq.\ \eqref{eq:norris6.2} becomes
\begin{align}
\ddot{p} &= \frac{K}{\sqrt{\bar \gamma}} \bar \partial_i \sqrt{\bar \gamma}\bar G^{ij} \bar \partial_j p\ .
\label{eq:norris11}
\end{align}
To conclude the program of transformation acoustics, the transformed wave equation should be re-interpreted in the original coordinates, which at the level of Eq.\ \eqref{eq:norris11} means that the barred partial derivatives should be replaced partial derivatives without a bar. However, to obtain the material parameters of the transformation acoustics device one has to get back to a wave equation in a covariant formulation,
\begin{align}
\ddot{p} &= \frac{\tilde K}{\sqrt{\gamma}} \partial_i \sqrt{\gamma}\tilde G^{ij} \ \partial_j p =\tilde K  D_i \tilde G^{ij} D_j p \ , & \tilde G^{ij} &= \tilde S^{ik} \tilde \rho^{-1}_{kl} \tilde S^{lj}\ .
\label{eq:norris12}
\end{align}
Equating  Eq.\ \eqref{eq:norris8} or \eqref{eq:norris11} with Eq.\ \eqref{eq:norris12} unveils the necessary material parameters of the transformation acoustics device:
\begin{align}
\tilde K &= \frac{\sqrt{\gamma}}{\sqrt{\bar{\gamma}}} K & \tilde S^{ij} \tilde \rho^{-1}_{jk} \tilde S^{kl} &= \frac{\sqrt{\bar \gamma}}{\sqrt{{\gamma}}} \bar S^{ij} \bar \rho^{-1}_{jk} \bar S^{kl}
\label{eq:norris13}
\end{align}
At this point, one has to remember the constraint \eqref{eq:norris2}, which played a crucial role in the derivation of the wave equation \eqref{eq:norris6}. Indeed, after the coordinate transformation $\bar{\bm S}$ satisfies the constraint $\bar D_i \bar S^{ij} = 0$, but in a transformation acoustic interpretation the altered material matrix has to satisfy $D_i \tilde S^{ij} = 0$. Therefore, the second equation in \eqref{eq:norris13} requires an appropriate redefinition of the elasticity tensor and the mass density matrix of the transformation acoustics material. As discussed in Ref.~\cite{Norris:2008Ps} this in general can only be achieved if an anisotropic mass density is assumed.

Though the above equations in principle define transformation acoustics for any pentamode material and any coordinate transformation, an important restriction has to be stressed: In contrast to transformation optics, the compatibility equations for the material parameters of the transformation medium are not entirely algebraic, but include the differential equation $D_i \tilde S^{ij} = 0$. The formalism of transformation acoustics does not guarantee that a solution to this equation can be found.

\section{Transformation acoustics for generic media}
\label{se:GenericMedia}
It is the aim of this section to show that a transformation scheme similar to Norris' theory of transformation acoustics conceptually can be defined for a much larger class of materials. In particular, we relax the constraint \eqref{eq:norris1} and in principle allow for an arbitrary set of eigenvalues $K_I$.

\subsection{The general divergence free material}
In a first step we derive a transformation acoustics scheme that follows closely the idea by Norris. Consequently, we assume that for a certain decomposition \eqref{eq:decomp} all $\bm S_I$ obey
\begin{equation}
D_i S_I^{ij}=0 \qquad \forall I\ .
\label{eq:Sdivfree}
\end{equation}
Obviously, this constrains the possible spatial dependence of the elasticity tensor in a quite drastic way. In Cartesian coordinates one simply concludes $D_i C^{ijkl} = \partial_i C^{ijkl} = \sum (\partial_i K_I)S_I^{ij} S_I^{kl}$.

The derivation of the wave equation closely follows the one of the previous section. For each matrix $\bm S_I$ a ``pressure'' can be defined as
\begin{equation}
p_I = - K_I S_I^{ij} D_i u_j.
\label{eq:pIdef}
\end{equation}
This allows to rewrite the stress tensor as $\sigma^{ij} = - \sum_I p_I S_I^{ij}$ and thus the momentum conservation becomes
\begin{equation}
D_j \sigma^{ij} = - \sum_I S_I^{ij} D_j p_I = \rho^i{}_j \ddot u^j\ .
\label{eq:divfree1}
\end{equation}
This equation can be solved for $\ddot{\bm u}$, the result is used in Eq.\ \eqref{eq:pIdef} twice differentiated with respect to time:
\begin{equation}
\begin{split}
\ddot p_I &= K_I S_{I}^{km} D_k \rho^{-1}_{mi} D_i \sum_J S_{J}^{ij} p_J \\
 &=  \sum_J K_I D_k S_{I}^{km} \rho^{-1}_{mi} S_{J}^{ij} D_j p_J\\
 &= \sum_J K_I D_i G^{ij}_{IJ} D_i p_J
\end{split}
\label{eq:WEQdivfree}
\end{equation}
As a result, the wave equation of elastodynamics has been transformed into six coupled wave equations of six scalars $p_I$. If the mass density matrix $\bm \rho$ is assumed as symmetric, these coupled wave equations are characterized by 21 effective metrics $\bm G_{IJ} = \bm G_{JI}$.

Despite the rather complicated form of the wave equations \eqref{eq:WEQdivfree}, this reformulation allows to proceed the transformation acoustic scheme along the lines of Norris' pentamode transformation acoustics. Applying the coordinate transformation $\bm x \longrightarrow \bar{\bm x}(x)$, Eq.\ \eqref{eq:WEQdivfree} becomes
\begin{align}
 \ddot p_I &= \sum_J K_I \bar D_i  \bar G_{IJ}^{ij} \bar D_j p_J = \sum_J \frac{K_i}{\sqrt{\bar \gamma}} \bar \partial_i \sqrt{\bar{\gamma}} \bar G_{IJ}^{ij} \bar \partial_j p_J\ .
\label{eq:WEQdivfreebar}
\end{align}
To reinterpret the result as mimicking of the coordinate transformation, the barred partial derivatives are again changed back to partial derivatives without bar. To obtain the correct covariant derivatives in the coordinate system $\bm x$, the known trick \eqref{eq:norris13} of rescaling by the square root of the metrics is used:
\begin{equation}
\begin{split}
 \ddot p_I &= \sum_J \frac{K_i}{\sqrt{\bar \gamma}} \partial_i \sqrt{\bar{\gamma}} \bar G_{IJ}^{ij} \partial_j p_J  \\
   &= \sum_J K_I \frac{\sqrt{\gamma}}{\sqrt{\bar\gamma}} D_k \left( \frac{\sqrt{\bar\gamma}}{\sqrt{\gamma}} \bar S_{I}^{km} \bar \rho^{-1}_{mi} \bar S_{J}^{ij} D_j p_J \right)\\
   &= \sum_J \tilde K_I D_i \left( \tilde S_{I}^{km} \tilde \rho^{-1}_{mi} \tilde S_{J}^{ij} D_j p_J \right)\ .
 \end{split}
\label{eq:WEQdivfreeTA}
\end{equation}
Thanks to the reformulation of the wave equation in terms of the scalars $p_I$, the rescaling method used by Norris (Eq.\ \eqref{eq:norris13}) also defines a transformation acoustics scheme for a generic divergence free material. However, the material parameters of the transformation medium are not defined in a unique way, but the possible realizations of the transformation medium are defined by the set of equations
\begin{align}
\tilde C^{ijkl} &= \sum_{I=1}^6 \tilde K_I \tilde S_I^{ij} \tilde S_I^{kl}\ ,\\
\label{eq:checkKdef}
\tilde K_I &= \frac{\sqrt{\gamma}}{\sqrt{\bar\gamma}} K_I\ ,\\
\label{eq:checkSdef}
\tilde S_I^{ik} \tilde \rho^{-1}_{kl} \tilde S_{J}^{lj} &= \frac{\sqrt{\bar\gamma}}{\sqrt{\gamma}} \bar S_I^{ik} \bar \rho^{-1}_{kl} \bar S_{J}^{lj}\ ,\\
\label{eq:checkSconstr}
D_i \tilde S_I^{ij} &= 0 \qquad \forall I\ .
\end{align}
Obviously, there exist solutions to the two algebraic constraints \eqref{eq:checkKdef} and \eqref{eq:checkSdef} for any choice of material parameters and coordinate transformation. However, as in the case of Norris' pentamode transformation acoustics, the last constraint is not an algebraic equation and it is not guaranteed that a solution exists. Still, a generic transformation acoustic scheme cannot be expected, since elastodynamics, in contrasts to electrodynamics, is not a premetric theory. Therefore, it is seen as a success of the reformulation in terms of the scalars $p_I$ that the consequences of the missing premetricity of the theory can be cast into the simple form of Eq.\ \eqref{eq:checkSconstr}. As will be shown below, the recipe derived in this section can be helpful in certain situations to obtain transformation acoustic results for materials different than just pentamode media.

\subsection{Decomposition of the general material}
Let's simply drop the condition that the matrices $\bm S_I$ should be divergence free. It was seen in the previous section that a reformulation of the wave equation in terms of the scalars $p_I$ was the important step to bring the tensorial wave equation of elastodynamics into a form that is accessible to the standard techniques of transformation acoustics. Since the definition of $p_I$ in Eq.\ \eqref{eq:pIdef} is purely algebraic, these scalars exist completely independently of the constraint \eqref{eq:Sdivfree}. However, the simple form of the momentum conservation as given in Eq.\ \eqref{eq:divfree1} is no longer correct; instead, that equation has to be replaced by
\begin{equation}
D_j \sigma^{ij} = - \sum_I D_j \left( S_I^{ij} p_I\right) = - \sum_I (j_I^i p_I + S_I^{ij} D_j p_I) = \rho^{i}{}_j \ddot u^j\ ,
\label{eq:general1}
\end{equation}
where we introduced the source terms $D_i S_I^{ij} = j^j_I$. From the definition of the scalars $p_I$ in Eq.\ \eqref{eq:pIdef} one obtains
\begin{equation}
\begin{split}
\ddot p_I &= - K_I S_I^{ij} D_i \ddot u_j = K_I S_I^{ij} D_i \left(\rho^{-1}_{jk} \sum_J D_l \left(S_J^{kl} p_J \right)\right)\\
&= \sum_J K_I \left(D_i S_I^{ij} - j_I^j\right) \rho^{-1}_{jk} \left( S_J^{kl} D_l + j_J^k\right) p_J\ . 
\end{split}
\label{eq:general2}
\end{equation}
As is seen, the reformulation in terms of the scalars $p_I$ still allows to reformulate the wave equation of elastodynamics in terms of six coupled differential equations in the six scalars $p_I$. While in the case of the divergence free material, these coupled equations where all of Helmholtz type, the equations \eqref{eq:general2} include additional inhomogeneous terms, which make the situation more complex. Despite this fact, one can proceed straightforwardly in the program of transformation acoustics. Since all quantities of Eq.\ \eqref{eq:general2} are just scalars, vectors or tensors the effect of a coordinate transformation $\bm x \longrightarrow \bar{\bm x}$ is merely to put a bar over each quantity. Since all covariant derivatives hit either a scalar or constitute the divergence of a vector, this barred equation again can be rewritten as
\begin{equation}
 \ddot p_I = \sum_J K_I \left(\frac{1}{\sqrt{\bar \gamma}}\bar \partial_i \sqrt{\bar \gamma} \bar S_I^{ij} - \bar j_I^j\right) \bar \rho^{-1}_{jk} \left( \bar S_J^{kl} \bar \partial_l + \bar j_J^k\right) p_J\ .
\label{eq:general3}
\end{equation}
The transformation acoustic scheme following from this transformation again interprets the barred partial derivatives as derivatives without bar. Transforming back to covariant derivatives without bar, the transformation acoustic analogue to Eq.\ \eqref{eq:general3} is
\begin{equation}
\begin{split}
 \ddot p_I &= \sum_J K_I \left(\frac{\sqrt{\gamma}}{\sqrt{\bar \gamma}} D_i \frac{\sqrt{\bar \gamma}}{\sqrt{\gamma}} \bar S_I^{ij} - \bar j_I^j\right) \bar \rho^{-1}_{jk} \left( \bar S_J^{kl} D_l + \bar j_J^k\right) p_J\\
 &= \sum_J \frac{\sqrt{\gamma}}{\sqrt{\bar \gamma}}K_I \left(D_i \bar S_I^{ij} - \bar j_I^j\right) \frac{\sqrt{\bar \gamma}}{\sqrt{\gamma}} \bar \rho^{-1}_{jk} \left( \bar S_J^{kl} D_l + \bar j_J^k\right) p_J \ .
\end{split}
\label{eq:general4}
\end{equation}
Comparison of the Eq.\ \eqref{eq:general4} with Eq.\ \eqref{eq:general2} unravels that the material parameters of an acoustic transformation material mimicking the coordinate transformation $\bm x \longrightarrow \bar{\bm x}$ follow from the set of equations
\begin{align}
\tilde C^{ijkl} &= \sum_{I=1}^6 \tilde K_I \tilde S_I^{ij} \tilde S_I^{kl}\ ,\\
\label{eq:general5}
\tilde K_I &= \frac{\sqrt{\gamma}}{\sqrt{\bar\gamma}} K_I\ ,\\
\label{eq:general6}
\tilde S_I^{ik} \tilde \rho^{-1}_{kl} \tilde S_{J}^{lj} &= \frac{\sqrt{\bar\gamma}}{\sqrt{\gamma}} \bar S_I^{ik} \bar \rho^{-1}_{kl} \bar S_{J}^{lj}\ ,\\
\label{eq:general7}
\tilde j^k_I \tilde \rho^{-1}_{kl} \tilde S_{J}^{lj} &= \frac{\sqrt{\bar\gamma}}{\sqrt{\gamma}} \bar j^k_I \bar \rho^{-1}_{kl} \bar S_{J}^{lj}\ ,\\
\label{eq:general8}
\tilde j^k_I \tilde \rho^{-1}_{kl} \tilde j_J^l &= \frac{\sqrt{\bar\gamma}}{\sqrt{\gamma}} \bar j^k_I \bar \rho^{-1}_{kl} \bar j_J^l\ ,\\
\label{eq:general9}
D_i \tilde S^{ij}_I &= \tilde j^j_I\ .
\end{align}
The new inhomogeneous terms yield two new algebraic conditions in Eqs.\ \eqref{eq:general7} and \eqref{eq:general8}. Still, solutions to all algebraic conditions \eqref{eq:general5}--\eqref{eq:general8} can be found for any choice of material parameters and coordinate transformation. More importantly, the inhomogeneity enters the differential definition of $\tilde j^j_I$ in Eq.\ \eqref{eq:general9}, which makes it very difficult in general to find a valid solution to all equations \eqref{eq:general5}--\eqref{eq:general9}.

An interesting simplification of the above equations occurs, if all $\bm S_I$ are either full rank or divergence free. This constitutes a restriction on the choice of material parameters, which however is less restrictive than the divergence free materials discussed in the previous section. In this particular case one can define the alternative source terms
\begin{equation}
 j_{I\, i} := \begin{cases} 0 & \text{if $D_i S_I^{ij} = 0$,}\\ S^{-1}_{I\, ij} D_k S_I^{kj} & \text{otherwise,} \end{cases}
\label{eq:fluxdef}
\end{equation}
and the wave equations \eqref{eq:general2} and \eqref{eq:general3} become
\begin{equation}
\begin{split}
   \ddot p_I &= \sum_J K_I (D_i-j_{I\, i}) S_{I}^{ik} \rho^{-1}_{kl} S_{J}^{lj} (D_j+j_{J\, j}) p_J\\
   &= \sum_J K_I (\frac{1}{\sqrt\gamma} \partial_i \sqrt{\gamma}-j_{I\, i}) S_{I}^{ik} \rho^{-1}_{kl} S_{J}^{lj}  (\partial_j+j_{J\, j}) p_J\\
   &= \sum_J \frac{K_I}{\sqrt{\gamma}} (\partial_i-j_{I\, i}) \sqrt{\gamma} G_{IJ}^{ij}  (\partial_j+j_{J\, j}) p_J \ ,
   \end{split}
\label{eq:WEQgeneral}
\end{equation}
where the effective metrics $\bm G_{IJ}$ are defined as in Eq.\ \eqref{eq:WEQdivfree}. The last equation differs from the respective form of the wave equation of the divergence free case only by the replacement $\bm \partial \longrightarrow \bm \partial \pm \bm j_I$. Evidently, the source terms do not affect the transformation acoustic interpretation of $\hat K_I$ and $\bm G_{IJ}$. Therefore, the material conditions of the acoustic metamaterial, which corresponds to the coordinate transformation $\bm x \longrightarrow \bar{\bm x}$, follow in analogy to the divergence free case, Eqs.\ \eqref{eq:checkKdef}--\eqref{eq:checkSconstr}, as
\begin{align}
\tilde C^{ijkl} &= \sum_{I=1}^6 \tilde K_I \tilde S_I^{ij} \tilde S_I^{kl}\ ,\\
\label{eq:gencheckKdef}
\tilde K_I &= \frac{\sqrt{\gamma}}{\sqrt{\bar\gamma}} K_I\ ,\\
\label{eq:gencheckSdef}
\tilde S_I^{ik} \tilde \rho^{-1}_{kl} \tilde S_{J}^{lj} &= \frac{\sqrt{\bar\gamma}}{\sqrt{\gamma}} \bar S_I^{ik} \bar \rho^{-1}_{kl} \bar S_{J}^{lj}\ ,\\
 D_i \tilde{S}^{ij} &= \begin{cases} 0 & \text{if $D_i S_I^{ij} = 0$,}\\ \bar j_{I\, i} \tilde{S}^{ij}  & \text{otherwise.} \end{cases}
\label{eq:Scheckjbar}
\end{align}
Since this formulation of material equations suppresses any dependence of the algebraic conditions on the source terms $\bm j_I$, it provides considerably more freedom to find solutions to these conditions that obey the definition of $\tilde j^j_I \equiv \bar j_{I\, i}$ from Eq.\ \eqref{eq:Scheckjbar}.

\section{Radial transformations and the cloak}
\label{se:examples}

In many applications the transformation device is either of spherical or cylindrical shape and one is interested in modifying the trajectories of waves exclusively in the radial direction. The best known examples are the spherical \cite{Pendry:2006Sc} and cylindrical \cite{Schurig:2006Sc} cloak. Thus, we want to investigate in this section if and how purely radial transformations can be extended by means of the formalism developed in this work. Various strategies for acoustic cloaks have been presented in the literature (see Ref.\ \cite{Chen:2010Ra} and references therein), but to our knowledge all these proposals assume an idealized situation where a single scalar wave is governed by a Helmholtz type equation. Since our formalism should be able to cover systems with more than one scalar mode it should be possible to find entirely new cloaking applications.

\subsection{Radial transformations in cylindrical coordinates}
Let us start by analyzing the situation of a cylindrical system. In cylindrical coordinates the metric tensor is given by
\begin{align}
[\gamma_{ij}] &= \mbox{diag}(1,r^2,1)\ , & [\gamma^{ij}] &= \mbox{diag}(1,\frac{1}{r^2},1)\ .
\label{eq:cyl1}
\end{align}
All information about the local structure of space and the choice of coordinates is encoded in this object. An important quantity in our formalism is the covariant divergence of the matrices $\bm S_I$, which is obtained from
\begin{equation}
D_i S^{ij}_I = \partial_i S^{ij}_I + \Gamma^i_{mi} S_I^{mj} + \Gamma^j_{mi} S_I^{im}\ .
\label{eq:cyl2}
\end{equation}
In cylindrical coordinates, the Christoffel symbols
\begin{equation}
\Gamma^i_{kl} = \frac{1}{2} \gamma^{im}\left(\partial_l \gamma_{mk} + \partial_k \gamma_{ml} - \partial_m \gamma_{kl} \right)
\label{eq:cyl3}
\end{equation}
are given by
\begin{align}
\bm \Gamma^r &= \mbox{diag}(0,-r,0)\ , & \bm \Gamma^\theta &= \begin{pmatrix}
	0&\frac1r&0\\\frac1r&0&0\\0&0&0
\end{pmatrix}\ , & \bm \Gamma^z = 0\ ,
\label{eq:cyl4}
\end{align}
which yields the divergence (remember that $\bm S_I$ must be symmetric)
\begin{align}
\label{eq:cyl5}
D_i S_I^{ir} &= \partial_i S_I^{ir} + \frac1r S_I^{rr} - r S_I^{\theta\theta}\ , \\
\label{eq:cyl6}
D_i S_I^{i\theta} &= \partial_i S_I^{i\theta} + \frac3r S_I^{r\theta}\ , \\
D_i S_I^{iz} &= \partial_i S_I^{iz} + \frac1r S_I^{rz}\ .
\label{eq:cyl7}
\end{align}

As has been worked out in the previous sections, the \emph{tour de force} of our transformation acoustics scheme is to relate the divergence of $\bm S_I$ after the coordinate transformation, $\bar D_i \bar S_I^{ij}$, to the divergence of a suitable choice of new material parameters $\tilde{\bm S}_I$, $D_i \tilde S_I^{ij}$. Therefore, we develop in a first step a strategy to solve this problem for radial transformations in cylindrical coordinates. Consider a linear radial transformation applied to our system,
\begin{align}
\bar r &= A r + B\ , & \bar \theta &= \theta\ , & \bar z &= z\ .
\label{eq:cylnew1}
\end{align}
Under such a transformation, the metric and the Christoffel symbols transform as
\begin{align}
\label{eq:cylnew2}
[\bar \gamma_{ij}] &= \mbox{diag}(\frac{1}{A^2},\frac{(\bar r-B)^2}{A^2} ,1)\ , & [\bar \gamma^{ij}] &= \mbox{diag}(A^2,\frac{A^2}{(\bar r-B)^2},1)\ , &\\
\bar{\bm \Gamma}^r &= \mbox{diag}(0,-(\bar r - B),0)\ , & \bar{\bm \Gamma}^\theta &= \begin{pmatrix}
	0&\frac1{\bar r-B}&0\\\frac1{\bar r-B}&0&0\\0&0&0
\end{pmatrix}\ , & \bar{\bm \Gamma}^z = 0\ .
\label{eq:cylnew3}
\end{align}
Therefore, the divergence of the matrices $\bar{\bm S}_I$ is given by
\begin{align}
\label{eq:cylnew4}
\bar D_i \bar S_I^{ir} &= \bar \partial_i \bar S_I^{ir} + \frac1{\bar r-B} \bar S_I^{rr} - (\bar r - B) \bar S_I^{\theta\theta}\ , \\
\label{eq:cylnew5}
\bar D_i \bar S_I^{i\theta} &= \bar \partial_i \bar S_I^{i\theta} + \frac3{\bar r-B} \bar S_I^{r\theta}\ , \\
\bar D_i \bar S_I^{iz} &= \bar \partial_i \bar S_I^{iz} + \frac1{\bar r-B} \bar S_I^{rz}\ .
\label{eq:cylnew6}
\end{align}
To solve the constraints \eqref{eq:checkSconstr}, \eqref{eq:general9} or \eqref{eq:Scheckjbar} new matrices $\tilde{\bm S}_I$ must be defined in such a way that $D_i \tilde S^{ij}$ can be related to $\bar D_i \bar S^{ij}$. Ideally, $D_i \tilde S^{ij}$ is chosen componentwise proportional to $\bar D_i \bar S^{ij}$:
\begin{equation}
D_i \tilde S^{ij} = \alpha^j \bar D_i \bar S^{ij} \qquad \text{no sum over j.}
\end{equation}
 In general, this will lead to a complicated system of differential equations.
In the example at hand, however, a simple solution can be found if we assume that all $\bar{\bm S}_I$ are diagonal in cylindrical coordinates. This means that we have at most three (linearly independent) matrices $\bm S_I$. The three independent $\bm S_I$ correspond to the three pressure modes, while the three shear modes become ``easy modes''. Since this material is shear free and in general has an anisotropic mass density as well as an anisotropic elasticity tensor, it can be seen as a generalization of the anisotropic metafluid \cite{Norris:2009Am}, for which a transformation acoustics scheme is known already.

With this simplifying assumption the covariant derivatives \eqref{eq:cylnew4}--\eqref{eq:cylnew6} become
\begin{align}
\label{eq:cylnew7}
\bar D_i \bar S_I^{ir} &= \bar \partial_r \bar S_I^{rr} + \frac1{\bar r-B} \bar S_I^{rr} - (\bar r - B) \bar S_I^{\theta\theta}\ , \\
\label{eq:cylnew8}
\bar D_i \bar S_I^{i\theta} &= \bar \partial_\theta \bar S_I^{\theta\theta}\ , \\
\bar D_i \bar S_I^{iz} &= \bar \partial_z \bar S_I^{zz}\ .
\label{eq:cylnew9}
\end{align}
The simple form of the covariant derivatives allows to define a linear transformation, $\tilde{\bm S}_I = \bm T \cdot \bar{\bm S}_I$, such that the divergence with respect to $\bar{\bm D}$ is transformed into a divergence with respect to $\bm D$. Indeed, the structure of the dependence in $\bar S^{rr}$ suggests the ansatz
\begin{equation}
\tilde{\bm S}_I = \exp[f(r)] \bar{\bm S}_I\ ,
\label{eq:cylnew10}
\end{equation}
where $f(r)$ follows from
\begin{equation}
\partial_r f(r) = \frac{1}{r-B} - \frac1r \qquad \Rightarrow \qquad f(r) = \ln \frac{r-B}{r}\ .
\label{eq:cylnew11}
\end{equation}
An additional scaling factor is needed to correct the $\bar S_I^{\theta\theta}$ term in \eqref{eq:cylnew7} and thus we find from the rescalings
\begin{align}
\tilde S_I^{rr} &= \frac{r-B}{r} \bar S_I^{rr}\ , & \tilde S_I^{\theta\theta} &=  \frac{(r-B)^2}{r^2} \bar S_I^{\theta\theta}\ , & \tilde S_I^{zz} &= \frac{r-B}{r} \bar S_I^{zz}\ ,
\label{eq:cylnew12}
\end{align}
the new covariant derivatives
\begin{align}
D_r \tilde S_I^{rr} &= \frac{r-B}{r} \bar D_r \bar S_I^{rr}\ , & D_\theta \tilde S_I^{\theta\theta} &=  \frac{(r-B)^2}{r^2} \bar D_\theta \bar S_I^{\theta\theta}\ , & D_z \tilde S_I^{zz} &= \frac{r-B}{r} \bar D_z \bar S_I^{zz}\ .
\label{eq:cylnew12.1}
\end{align}
The rescaling of the $S^{zz}_I$ components is not fixed by Eqs.\ \eqref{eq:cylnew7}--\eqref{eq:cylnew9} but has been chosen for convenience.

Since all matrices $\bar{\bm S}_I$ are diagonal, the above rescalings indeed may written as the matrix equations
\begin{align}
\tilde{\bm S}_I &= \bm T \cdot \bar{\bm S}_I = \bar{\bm S}_I \cdot \bm T\ , & \bm T &= \mbox{diag}(\frac{r-B}{r},\frac{(r-B)^2}{r^2},\frac{r-B}{r})\ .
\label{eq:cyl18}
\end{align}
Most importantly, the matrix $\bm T$ has been constructed in such a way that
\begin{equation}
D_i T^i{}_j \bar S^{jk} = T^k{}_j \bar D_i \bar S^{ij}\ .
\label{eq:cylnew15}
\end{equation}

The construction performed so far solves the differential constraint \eqref{eq:general9}. From this result, the remaining constraints Eqs.\ \eqref{eq:general5}--\eqref{eq:general8} are easily solved. Indeed, from the ``commutation relation'' \eqref{eq:cylnew15} it follows that $\tilde j^j_I$ in Eq.\ \eqref{eq:general9} is obtained by the same rescaling matrix as $\tilde S^{ij}$:
\begin{equation}
\tilde j^j_I = T^j{}_i \bar j^i_I
\label{eq:cylnew16}
\end{equation}
This, in turn, implies that on the left hand sides of Eqs.\ \eqref{eq:general6}--\eqref{eq:general8} the same resaclings are applied and thus these three equations are simultaneously solved by
\begin{equation}
\tilde{\bm \rho} = \frac{\sqrt{\gamma}}{\sqrt{\bar{\gamma}}} \bm T \cdot \bar{\bm \rho} \cdot \bm T\ .
\label{eq:cylnew17}
\end{equation}
Eqs.\ \eqref{eq:cyl18} and \eqref{eq:cylnew17} define the material parameters of a transformation acoustics scheme for a coordinate transformation of type \eqref{eq:cylnew1} and for a material with three eigentensors $\bm S_I$. These tensors have to be diagonal in cylindrical coordinates but need not be of full rank or divergence free.

\subsection{Radial transformations in spherical coordinates}
In spherical coordinates $(r,\theta,\phi)$ the metric tensor reads
\begin{align}
[\gamma_{ij}] &= \mbox{diag}(1,r^2 \sin^2 \phi,r^2)\ , & [\gamma^{ij}] &= \mbox{diag}(1,\frac{1}{r^2 \sin^2\phi},\frac{1}{r^2})\ ,
\label{eq:sph1}
\end{align}
which yields the Christoffel symbols
\begin{align}
\bm \Gamma^r &= \mbox{diag}(0,-r \sin^2 \phi,-r)\ , & \bm \Gamma^\theta &= \begin{pmatrix}
	0&\frac1r&0\\\frac1r&0&\cot \phi\\0&\cot \phi&0
\end{pmatrix}\ , & \bm \Gamma^z = \begin{pmatrix}
	0&0&\frac1r\\0&-\sin\phi\cos\phi&0\\\frac1r&0&0
\end{pmatrix}\ .
\label{eq:sph2}
\end{align}
Thus, the divergence of the matrices $\bm S_I$ in these coordinates reads
\begin{align}
\label{eq:sph3}
D_i S_I^{ir} &= \partial_i S_I^{ir} + \frac2r S_I^{rr} + \cot \phi S_I^{\phi r} - r \sin^2\phi S_I^{\theta\theta} - r S_I^{\phi\phi}\ , \\
\label{eq:sph4}
D_i S_I^{i\theta} &= \partial_i S_I^{i\theta} + \frac4r S_I^{r\theta} + 3 \cot\phi S_I^{\phi\theta}\ , \\
D_i S_I^{i\phi} &= \partial_i S_I^{i\phi} + \frac4r S_I^{r\phi} + \cot\phi S_I^{\phi\phi}-\sin\phi\cos\phi S_I^{\theta\theta}\ .
\label{eq:sph5}
\end{align}
In analogy to the previous section we apply a linear radial transformation
\begin{align}
\bar r &= A r + B\ , & \bar \theta &= \theta\ , & \bar \phi &= \phi\ .
\label{eq:sph6}
\end{align}
The metric tensor under this transformation changes to
\begin{align}
[\bar \gamma_{ij}] &= \mbox{diag}(\frac{1}{A^2},\frac{(\bar r-B)^2}{A^2} \sin^2 \phi,\frac{(\bar r-B)^2}{A^2})\ , & [\bar \gamma^{ij}] &= \mbox{diag}(A^2,\frac{A^2}{(\bar r-B)^2 \sin^2\phi},\frac{A^2}{(\bar r-B)^2})\ ,
\label{eq:sph7}
\end{align}
As in cylindrical coordinates, all factors $A^2$ drop out in the calculation of the Christoffel symbols, such that they are obtained from the expressions in Eq.\ \eqref{eq:sph2} simply by replacing $r$ everywhere by $\bar r - B$. Consequently, the divergence of $\bar{\bm S}_I$ in spherical coordinates is given by
\begin{align}
\label{eq:sph8}
\bar D_i \bar S_I^{ir} &= \bar \partial_i \bar S_I^{ir} + \frac2{\bar r-B} \bar S_I^{rr} + \cot \bar\phi \bar S_I^{\phi r} - (\bar r-B) \sin^2\bar \phi \bar S_I^{\theta\theta} - (\bar r-B) \bar S_I^{\phi\phi}\ , \\
\label{eq:sph9}
\bar D_i \bar S_I^{i\theta} &= \bar\partial_i \bar S_I^{i\theta} + \frac4{\bar r-B} \bar S_I^{r\theta} + 3 \cot\bar\phi \bar S_I^{\phi\theta}\ , \\
\bar D_i \bar S_I^{i\phi} &= \bar \partial_i \bar S_I^{i\phi} + \frac4{\bar r-B} \bar S_I^{r\phi} + \cot\bar\phi \bar S_I^{\phi\phi}-\sin\bar\phi\cos\bar\phi \bar S_I^{\theta\theta}\ .
\label{eq:sph10}
\end{align}
In order to reinterpret this result in the original coordinate system $\bm x$ the same strategy as in the previous case proves successful: First, we demand that all $\bar{\bm S}_I$ are diagonal in spherical coordinates, which simplifies Eqs.\ \eqref{eq:sph8}--\eqref{eq:sph10} to
\begin{align}
\label{eq:sph11}
\bar D_i \bar S_I^{ir} &= \bar \partial_r \bar S_I^{rr} + \frac2{\bar r-B} \bar S_I^{rr}  - (\bar r-B) \sin^2\bar \phi \bar S_I^{\theta\theta} - (\bar r-B) \bar S_I^{\phi\phi}\ , \\
\label{eq:sph12}
\bar D_i \bar S_I^{i\theta} &= \bar\partial_\theta \bar S_I^{\theta\theta}\ , \\
\bar D_i \bar S_I^{i\phi} &= \bar \partial_\phi \bar S_I^{\phi\phi} + \cot\bar\phi \bar S_I^{\phi\phi}-\sin\bar\phi\cos\bar\phi \bar S_I^{\theta\theta}\ .
\label{eq:sph13}
\end{align}
Again, a diagonal matrix can be defined, which relates between the divergence with respect to $\bar{\bm \gamma}$ and with respect to $\bm \gamma$. Instead of the rescaling matrix $\bm T$ in Eq.\ \eqref{eq:cyl18}, a slightly different matrix $\bm U$
\begin{align}
\tilde{\bm S}_I &= \bm U \cdot \bar{\bm S}_I = \bar{\bm S}_I \cdot \bm U\ , & \bm U &= \mbox{diag}(\frac{r-B}{r},\frac{(r-B)^2}{r^2},\frac{(r-B)^2}{r^2})\ .
\label{eq:sph14}
\end{align}
must be used here, which also fulfills the important ``communtation relation'' \eqref{eq:cylnew15}. Therefore the material parameters in case of a spherical transformation are obtained from Eq.\ \eqref{eq:sph14} and
\begin{align}
\tilde j^j_I &= U^j{}_i \bar j^i_I\ , &
\tilde{\bm \rho} &= \frac{\sqrt{\gamma}}{\sqrt{\bar{\gamma}}} \bm U \cdot \bar{\bm \rho} \cdot \bm U\ .
\label{eq:sph15}
\end{align}

\section{Summary and conclusions}
In this work we developed a generic theory of transformation acoustics applicable to any elasticity tensor. In contrast to electromagnetics, elastodynamics is not a premetric theory and thus it is not expected that a transformation based design scheme exists for any choice of material parameters and coordinate transformations. Our approach to attack this problem is based on the previous works by Milton et al.\ \cite{Milton:2006Ac} and Norris \cite{Norris:2008Ps}. Following these authors we reformulate the wave equation in terms of the moduli $K_I$ and the eigentensors $\bm S_I$ of the elasticity tensor. In this formulation the complex tensorial wave equation turns into six coupled wave equations in six scalar quantities $p_I$. In consequence, for each of these equations a transformation scheme similar to the one of the Helmholtz equation can be invoked. As a result, compatibility equations for the eigentensors and moduli are found according to which the transformation material has to be defined.

Of course, the metric dependence of the theory is preserved by this reformulation and thus the related problems with transformation acoustics cannot be eliminated. In our formulation, these limitations result in non-algebraic compatibility equations (see Eqs.\ \eqref{eq:checkKdef}--\eqref{eq:checkSconstr}, Eqs.\ \eqref{eq:general5}--\eqref{eq:general9} and Eqs.\ \eqref{eq:gencheckKdef}--\eqref{eq:Scheckjbar}). Still, as a main result of this article we find that thanks to our reformulation the complicated dependence of the tensorial wave equation on the spatial metric can be hidden completely and that the metric dependence can be reduced to simple conditions on the divergence of the six eigentensors $\bm S_I$. To our knowledge this scheme for the first time provides a realistic transformation tool outside of electromagnetics that is applicable to more than just a single scalar mode.

As an application of our algorithm we studied linear radial transformations in cylindrical and spherical coordinates, which include the cylindrical and spherical cloak. It was shown that in this special case transformation media can be found for any material with three pressure modes but vanishing shear modes (anisotropic liquid). This result constitutes a generalization of previous studies of liquids in transformation acoustics, where the anisotropy of the metafluid only concerned its mass density tensor \cite{Norris:2009Am}.

Our result suggests that the transformational design approach could be a powerful tool in a much broader context than known today. The basis for such applications seems to be a clever reformulation of the wave equation in such a way that the required coordinate transformation easily can be implemented. A similar result has recently been developed by Garcia-Meca et al.\ \cite{Garcia-MecaIP}: these authors show, though for a different system than in this article, that with a suitable choice of variables certain transformation acoustics schemes become possible or exhibit different solutions than known before.
Still, the results found so far may be insufficient in various applications. While in electromagnetics transformation devices can be matched without any reflections to the surrounding space \cite{Bergamin:2009In}, interface conditions between standard elastic media and acoustic metamaterials from transformation acoustics have not been studied extensively so far.

\subsection*{Acknowledgement}
The author would like to thank Alberto Favaro for extensive discussions on this topic. Also, he is indebted to Martin McCall and Sante Carloni for useful comments. This work was supported in parts by the European Space Agency (ESA) via the Ariadna initiative of the Advanced Concepts Team (contract number 11-1301-a).

\bibliographystyle{fullsort}
\bibliography{bibliomaster}

\end{document}